\documentclass[10pt,a4paper]{article}
\usepackage[marginparwidth=2in]{geometry}%top=0.85in,footskip=0.75in,

\usepackage[utf8]{inputenc}
\usepackage{lmodern}

\usepackage{cite}

\usepackage{nameref,hyperref}

\usepackage{orcidlink}

\usepackage[left]{lineno}

\usepackage{microtype}
\DisableLigatures[f]{encoding = *, family = * }

\usepackage{changepage}

\usepackage[aboveskip=1pt,labelfont=bf,labelsep=period,singlelinecheck=off]{caption}

\makeatletter
\renewcommand{\@biblabel}[1]{\quad#1.}
\makeatother

\usepackage{lastpage,fancyhdr,graphicx}
\usepackage{epstopdf}
\usepackage{color}

\definecolor{Gray}{gray}{.25}

\usepackage{nicematrix}

\begin{document}
	\vspace*{0.35in}
	
	% title goes here:
	\begin{flushleft}
		{\Large
			\textbf\newline{Towards a sensing model using random laser combined with diffuse reflectance spectroscopy}
		}
		\newline
		% authors go here: 
		\\
		Dongqin Ni\textsuperscript{1,2,*} \orcidlink{0000-0003-2533-8932},
		Florian Kl\"ampfl\textsuperscript{1,2},
		Michael Schmidt\textsuperscript{1,2},
		Martin Hohmann\textsuperscript{1,2}\orcidlink{0000-0001-5099-1015}
		\\
		\bigskip
		\bf{1} Institute of Photonic Technologies (LPT), Friedrich-Alexander-Universit\"at Erlangen-N\"urnberg (FAU), Konrad-Zuse-Stra\ss e~3/5, 91052 Erlangen, Germany
		\\
		\bf{2} Erlangen Graduate School in Advanced Optical Technologies (SAOT), Paul-Gordan-Stra\ss e~6, 91052 Erlangen, Germany
		\\
		\bigskip
		* Dongqin.Ni@FAU.de
		
	\end{flushleft}
	
	\section*{Abstract}
	The previous research proves that the random laser emission reflects not only the scattering properties but also the absorption properties. The random laser is therefore considered a potential tool for optical properties sensing. Although the qualitative sensing using the random laser is extensively investigated, a quantitative measurement is still rare. In this study, a generalized mathematical quantitative model using random laser combined with diffuse reflectance spectroscopy is proposed for optical sensing in turbid media. This model describes the gain effect of the active medium and the optical properties effect of the passive medium separately. Rhodamine 6G is used as the active medium. Intralipid and ink are employed to demonstrate the effect of the scattering and absorption, respectively. The peak wavelength shift of the random laser is proved to be an ideal sensing parameter for this sensing model. It is also revealed that the scaling parameters in the sensing model are interrelated and can be simplified to one. With this combined model, the direct sensing of optical properties in diverse turbid media is promising. 
	
	%\linenumbers
	\section{Introduction}
	The understanding of light-matter interactions and the characterization of optical properties (OPs) are of paramount importance in various fields of research~\cite{Flory2011,Jacques2013}. In biological and biomedical fields, a profound understanding of OPs is essential for the interpretation of diagnostic and therapeutic measurements, as well as for the development of medical technologies~\cite{Jacques2013}. The characterization of OPs in turbid media, such as biological tissues, typically involves parameters such as absorption coefficient $\mu_a$, reduced scattering coefficient $\mu_s'$, and anisotropy factor $g$~\cite{Jacques2013}. While absorption and scattering strength are described by $\mu_a$ and $\mu_s'$, respectively, $g$ denotes the scattering angle dependence~\cite{Jacques2013}. The relation between the scattering parameters of $\mu_s'$ and $g$ is expressed by the equation:
	\begin{equation}
		\mu_s'=\mu_s(1-g),
		\label{eq_OPs}
	\end{equation}
	where $\mu_s$ represents the scattering coefficient. 
	
	The use of a random laser (RL) is emerging as a promising approach for the direct characterization of OPs in turbid media~\cite{Hohmann2021}. Within a RL system, the optical feedback for lasing stems from multiple scattering among random scatterers~\cite{Ni2023}. Consequently, the emission properties of RLs reflect the scattering properties of the random media, offering potential applicability in optical sensing in turbid media, particularly in biological tissues~\cite{Ni2023}. For example, RLs have demonstrated success in tissue differentiation~\cite{Hohmann2019}, cancerous tissue screening~\cite{Polson2004,Mogharari2019}, and the detection of optomechanical strain in tissues~\cite{Song2010}. These achievements are noteworthy, although the aforementioned sensing applications are found solely on qualitative measurements, thereby limiting their generalizability to diverse turbid systems. There is a need for a quantitative model using RL emission properties to assess the universal OPs in various turbid media. 
	
	Researchers have attempted to formulate a quantitative OPs sensing model based on RL emission~\cite{TOMMASI2018,Hohmann2021}. The sensing parameters include the RL emission intensity, peak wavelength, linewidth, lasing modes and lasing threshold~\cite{Ni2023}. In particular, the work of Tommasi et al.~\cite{TOMMASI2018} demonstrated the measurement of RL emission intensity to characterize the constant value of $\mu_s'$ arising from microspheres with different diameters. They separated the active gain medium from the passive scattering medium by using an isolated transparent spherical cell. Both pump and emission light were guided through one fiber. In the work from Hohmann et al.~\cite{Hohmann2021}, the active gain medium was mixed with the scattering medium. They reported the RL dependence on changes in $\mu_s$, not only from the RL intensity, but also from the spectral peak wavelength and linewidth. The responses to changes in $\mu_s$ are consistent for all three parameters. Furthermore, it was empirically found that critical alterations in the RL emission variations are induced by $\mu_s$ rather than $\mu_s'$. Specifically, the RL emission exhibited maximum intensity when the laser cavity length was an integer multiple of the scattering mean free path $l_s$ (equivalent to $1/\mu_s$), rather than the reduced scattering mean free path $l_s'$ (equivalent to $1/\mu_s'$)~\cite{Hohmann2021}. Consequently, direct measurement employing RLs may provide a solution to the precise characterization of $\mu_s$. The question remains in the optical characterization of $\mu_a$. It was explored in a previous study~\cite{Ni2023a}, wherein an increase in $\mu_a$ resulted in a decrease in RL intensity, a broadening of the linewidth, a blueshift of the peak wavelength, and an increase in the lasing threshold~\cite{Ni2023a}. It has been demonstrated that the RL emission exhibits an inverse behavior on $\mu_a$ in comparison to $\mu_s$ or $\mu_s'$~\cite{Ni2023a}. The above three studies showed the feasibility of the RL emission to sense the OPs of $\mu_s'$, $\mu_s$ and $\mu_a$. However, a mathematical representation of the sensing model is still lacking.
	
	In fact, the aforementioned RL behavior is analogous to the well-known diffuse reflectance: diffuse reflectance intensity is directly proportional to $\mu_s'$ and inversely proportional to $\mu_a$~\cite{Moy2016}. The phenomenon that the RL responds similarly to the diffuse reflectance in the context of OPs is not fortuitous. Pioneering RL research has evidenced that RL generated from a turbid medium with scattering strength in the diffuse regime can be modeled as light diffusion with gain~\cite{Wiersma1996}. In this study, when observed in the backscattered direction, the RL can be hypothesized to be light diffuse reflectance with gain. Since the modeling of the diffuse reflectance spectroscopy (DRS) signal is well established in literature~\cite{Moy2016,Akter2018}, the RL emission has high potential to be also modeled and applied for optical sensing. 
	
	In addition, the utilization of RL emission for sensing compared to the diffuse reflectance alone may facilitate deeper detection, since the typical sensing depth using diffuse reflectance is limited, for example, 0.5 - 1.9~mm in depth using a light source operating in the wavelength range of 350 - 1919~nm and applying a source-detector separation of 2500~{\textmu}m~\cite{Nogueira2021}. This limitation will be overcome in RLs by cascaded gain amplification along the light path, leading to an extended active light path length in turbid media~\cite{Ni2023}. Furthermore, this effect may improve the sensitivity of RL-based OPs sensing due to the enhanced laser emission compared to the lamp typically used in the diffuse reflectance measurements. Therefore, the prospect of constructing a mathematical RL-based OPs sensing model, adapted from the diffuse reflectance model, with a higher sensitivity of a deeper sensing target, seems very promising. 
	
	The aim of this study is therefore to use the RL spectral properties to construct a mathematical DRS-RL model for quantitative OPs sensing in turbid media. In particular, for the first time, this RL model is able to separately describe the gain effect in addition to the OPs effect. This model can be adapted to various turbid media, but in this study it is validated only in the simplest turbid system consisting of Intralipid (IL) as scatterer and black ink as absorber. 
	
	\section{Theoretical consideration}
	\subsection{DRS model}
	The diffuse reflectance is indeed a function of $\mu_s'$, $\mu_a$, phase function $p(\theta)$ (or $g$ factor in the multiple scattering regime in this study) and geometry $G$~\cite{Calabro2014a}:
	\begin{equation}
		R_d=f(\mu_s', \mu_a, g, G). 
		\label{eq_DR_model}
	\end{equation}
    The phase function $p(\theta)$ describes the scattering angle dependence in a single scattering event, analogous to the function of the $g$ factor in the multiple scattering condition. One study demonstrated the non-negligible influence of the phase function $p(\theta)$ on the diffuse reflectance in certain probe geometry configurations, e.g. when the light collection is close to the source~\cite{Kanick2012}. The probe geometry $G$ includes the effect of the source-detector separation and the effect of the numerical aperture of the fiber probe~\cite{Zonios2006,Hennessy2014}. The symbol $f()$ denotes a nonspecific function for the argument in bracket. The same function notation is applied in the following equations in this study.
	
	Among various applied mathematical models~\cite{Zonios2006}, the Zonios's model of diffuse reflectance in semi-infinite turbid media with fiber probes is the most practical one~\cite{Zonios2006}. The diffuse reflectance $R_d$ in the Zonios's model reads:
	\begin{equation}
		R_d=\frac{\mu_s'}{k_1+k_2\mu_a}, 
		\label{eq_DR}
	\end{equation}
	where $k_1$ and $k_2$ are the scaling parameters depending on the optical probe geometry $G$~\cite{Zonios2006}. The phase function was not formulated in this model.
	
	Nevertheless, Equation~\ref{eq_DR_model} and~\ref{eq_DR} indicate that a direct measurement of the OPs of $\mu_s'$, $\mu_s$ and $\mu_a$ is promising ($\mu_s$ can be calculated with known $\mu_s'$ and $g$ according to the Equation~\ref{eq_OPs}). The difficulty lies in the measurement of $g$. Although $g$ is a relevant parameter of OPs, a mathematical representation that separates its effect from the scaling parameters $k_1$ and $k_2$ in Equation~\ref{eq_DR} is still lacking in the DRS literature. 
	
	\subsection{DRS-RL sensing principle: sensing parameter and sensing curve}
\begin{figure}[h!]
		\centering
		\includegraphics[width=1\textwidth]{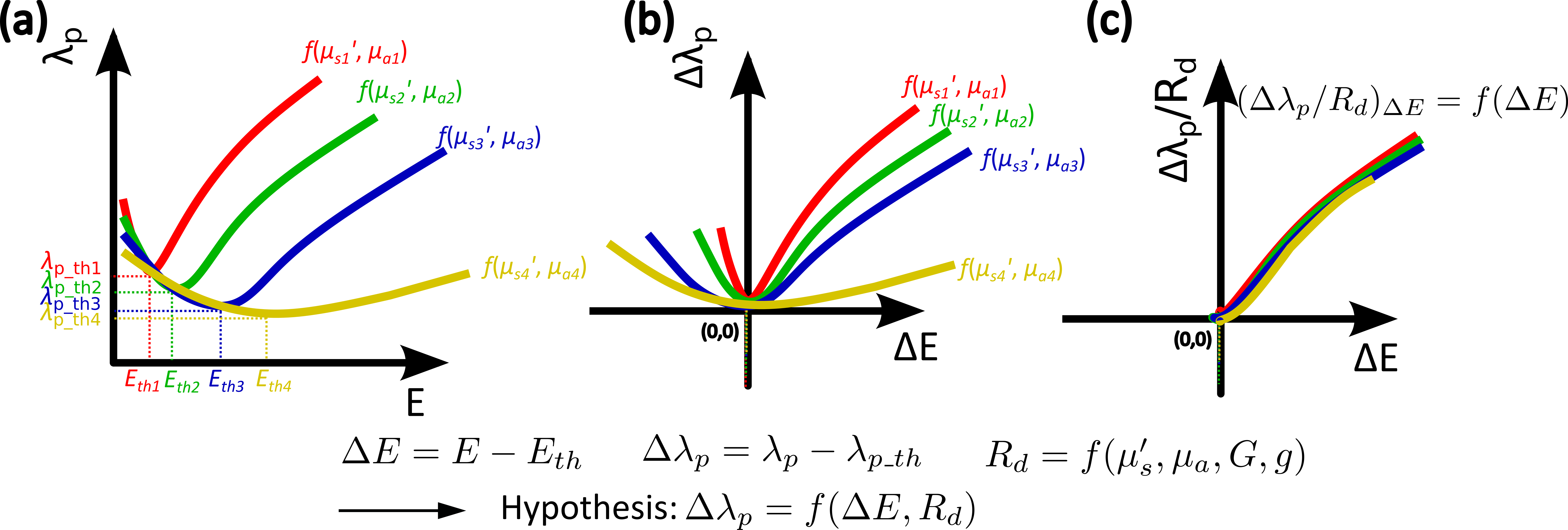}
		\caption{Sensing principle: (a) raw data showing the response of the RL peak wavelength $\lambda_p$ on the pump energy $E$. The lasing threshold is determined upon the peak wavelength changes from blueshift to redshift. Different colors represent the samples with different OPs of $\mu_s'$ and $\mu_a$. (b) pre-processed data to separate the non-lasing and lasing regime. $\lambda_p$ and $E$ are replaced by the relative peak wavelength $\Delta\lambda_p$ and the relative pump energy $\Delta E$, both taking the values at the lasing threshold as a reference. Only the data representing the lasing emission on the non-negative axes are further utilized for sensing. (c) converged sensing curve. The $\Delta \lambda_p$ of all samples after the correction of the $R_d$ effect, i.e. $\Delta \lambda_p / R_d$, responds identically to $\Delta E$, due to the fact that the remaining gain effect is identical in all samples.} 
		\label{fig_Hypothesis}
	\end{figure}

	The emission from the RL sample is either non-lasing or lasing emission. Only the RL signal in the lasing regime is utilized for sensing, because the RL emission in the lasing regime is more stable and stronger due to the dominant stimulated emission. To separate the non-lasing and lasing emissions, the peak wavelength of the RL is employed. As sketched in Figure~\ref{fig_Hypothesis}~(a), the peak wavelength $\lambda_p$ always changes from blueshift to redshift at the lasing threshold when the pump energy $E$ is increased. This changed trend of the peak wavelength shift at the lasing threshold is universal and independent of the OPs of the samples~\cite{Ni2023a}. Moreover, this non-monotonic change of the peak wavelength makes it a better lasing indicator than the other RL parameters which only show monotonic changes at the lasing threshold~\cite{Ni2023a}. Hence, the RL peak wavelength is chosen as the sensing parameter of the model. 
	
	In practice, to separate the non-lasing and lasing regimes, the peak wavelength and pump energy are replaced by a relative peak wavelength $\Delta\lambda_p=\lambda-\lambda_{th}$ and a relative pump energy $\Delta E=E-E_{th}$, both of which take the values at the lasing threshold as a reference. In other words, both $\Delta\lambda_p$ and $\Delta E$ are zero at the lasing threshold, as shown in Figure~\ref{fig_Hypothesis}~(b). Only the lasing regime where the $\Delta\lambda_p$ and $\Delta E$ are non-negative is used for the further sensing application. Meanwhile, the nonlinearity of the lasing effect at the threshold is also excluded by using $\Delta\lambda_p$ instead of $\lambda_p$ as the sensing parameter, which makes the formulation of the gain effect much easier. 
	
\begin{figure}[h!]
		\centering
		\includegraphics[width=0.6\textwidth]{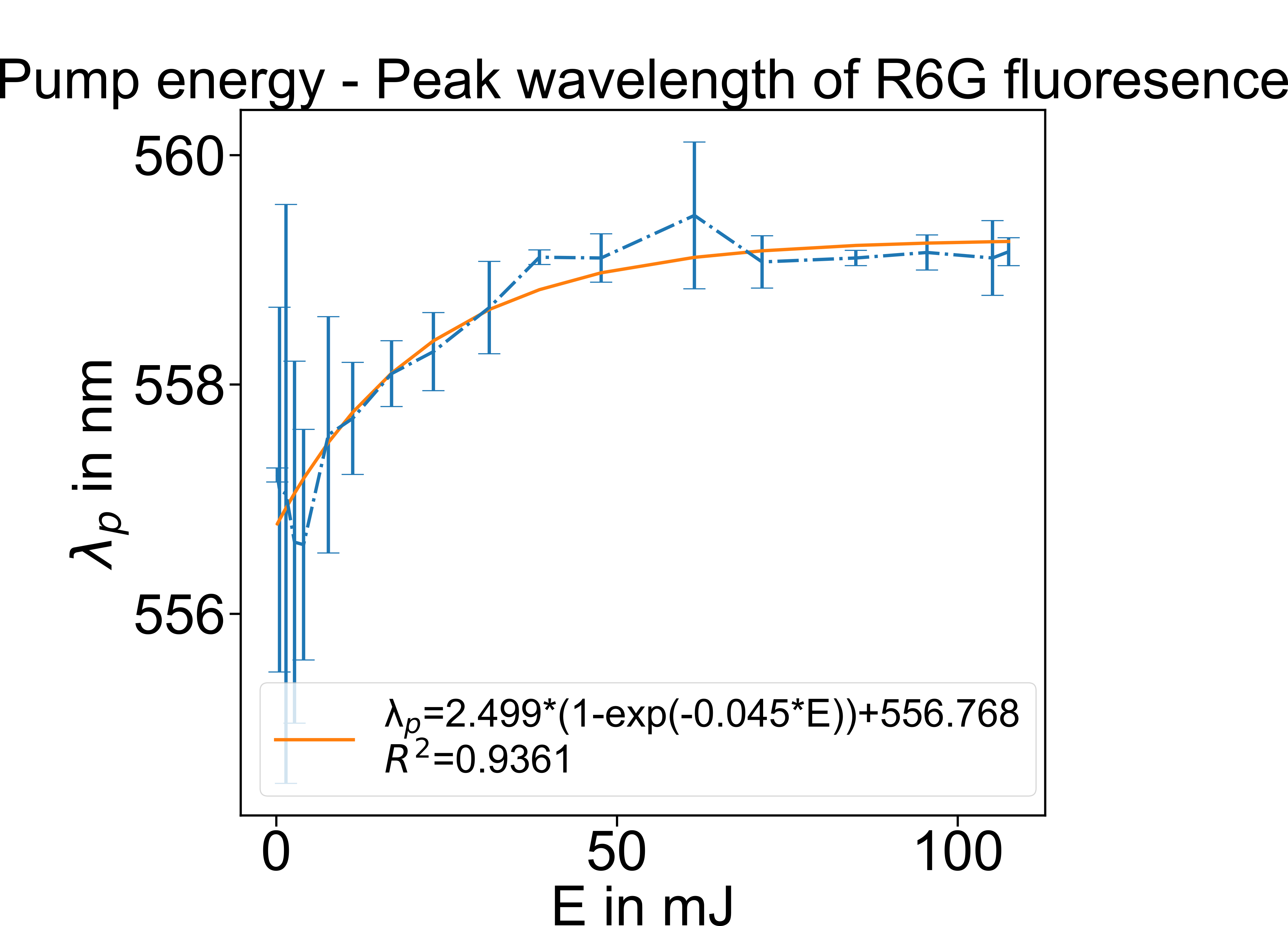}
		\caption{Response of peak wavelength to the pump energy applied to a R6G gain medium.} 
		\label{fig_Pure_R6G}
	\end{figure}

	The hypothesis that the RL emission is the light diffuse reflectance with gain is then mathematically expressed as:
	\begin{equation}
		\Delta \lambda_p = f(R_d,\Delta E).
		\label{eq_hypothesis}
	\end{equation}
	Since the response of the peak wavelength to the pump energy is OPs or diffuse reflectance independent, the gain effect induced by $\Delta E$, i.e. $f(\Delta E)$ can be formulated separately: 
	\begin{equation}
		\Delta \lambda_p = R_d \cdot f(\Delta E).
		\label{eq_hypothesis0}
	\end{equation}
	When the effect of $R_d$ is excluded from $\Delta\lambda_p$ by division, it is assumed that the $\Delta \lambda_p / R_d$ depends only on the $\Delta E$, i.e. the gain effect. Mathematically, this relationship can be expressed as:
	\begin{equation}
		\Delta \lambda_p / R_d = f(\Delta E)
		\label{eq_hypothesis1}
	\end{equation}
	and by applying the Zonios's model from Equation~\ref{eq_DR}: 
	\begin{equation}
		\Delta \lambda_p \cdot \frac{k_1+k_2\mu_a}{\mu_s'} = f(\Delta E).
		\label{eq_hypothesis2}
	\end{equation}
	The right side of the Equation~\ref{eq_hypothesis1} and~\ref{eq_hypothesis2} can be interpreted as the "OPs-independent gain effect" that can be measured when the sample has no scatterer and no absorber, but only a gain medium, i.e. a transparent active medium. In this study, the measurement of the "OPs-independent gain effect" was performed on the active medium of the Rhodamine 6G (R6G) water solution, as shown in Figure~\ref{fig_Pure_R6G}.	
	
	Since the gain effect of the same gain medium is identical, the response of $\Delta \lambda_p / R_d$ to $\Delta E$ is also identical for each sample. This identical response is sketched in Figure~\ref{fig_Hypothesis}~(c), where all RL peak wavelength after correcting for the OPs effect converge to an identical curve. This curve is referred to the "OPs-independent sensing curve" in this study. 
	
	Since the "OPs-independent gain effect" shows an exponential response in Figure~\ref{fig_Pure_R6G}, the fit function of the "OPs-independent sensing curve" is predefined as an exponential function too:
	\begin{equation}
		\Delta \lambda_p \cdot \frac{k_{1init}+k_{2init}\mu_a}{\mu_s'} = -a \cdot e^{-b \cdot \Delta E}+c,
		\label{eq_fitting}
	\end{equation}
	where $a$, $b$, $c$ are the fitting parameters, and $k_{1init}$, $k_{2init}$ are the initial assumptions of the scaling parameters of $k_1$, $k_2$. 
	
	In a short summary, the RL emission detected from the backscattered direction is assumed to be the diffuse reflectance with gain. The RL peak wavelength shift after correcting for the diffuse reflectance effect is assumed to be identical due to the identical gain medium. Such a response can be expressed mathematically as Equation~\ref{eq_fitting} and illustrated graphically as the identical sensing curve in Figure~\ref{fig_Hypothesis}~(c). Given the sensing curve, the measured $\Delta \lambda_p$ and $\Delta E$, and the optimized scaling parameters $k_1$ and $k_2$, the OPs of $\mu_s'$ and $\mu_a$ can be derived from Equation~\ref{eq_fitting}.
	
	\section{Experimental validation} 
	\subsection{RL samples and experimental setup}
	The RL sample prepared in this study is a water-based liquid medium, consisting of Rhodamine 6G (R6G; Sigma Aldrich, Germany) as the laser gain and Intralipid (IL; Fresenius Kabi, Germany) as the scatterer. Indian black ink (Royal Talens, The Netherland) was added as an external absorber. While the IL and ink concentrations were varied to change the scattering and absorption strength of the turbid media, the R6G concentration was kept at 2×10$^{-4}$~g/ml to achieve the optimal gain efficiency. 
	
	The method for calculating the OPs of scatters and absorbers was described in the previous study~\cite{Ni2023a,Aernouts2014}. Two concentrations of IL scatterers of 5\% and 9\%~v/v were selected, leading to $\mu_s'$ of 63.78 and 106.13~cm$^{-1}$, respectively. Since $\mu_s'$ is wavelength dependent, the above values took the average of the individual values calculated at different peak wavelengths of the RL. The ink concentrations ranged from 0 to 0.09\%~v/v, corresponding to $\mu_a$ from 0 to 8.64~cm$^{-1}$. The OPs values of $\mu_s'$ and $\mu_a$ were selected according to the OPs of biological tissues~\cite{Jacques2013}. Meanwhile, the values satisfy the diffusion approximation where $\mu_s'$ is much larger than $\mu_a$. 
	
	\begin{figure}[h!]
		\centering
		\includegraphics[width=0.6\textwidth]{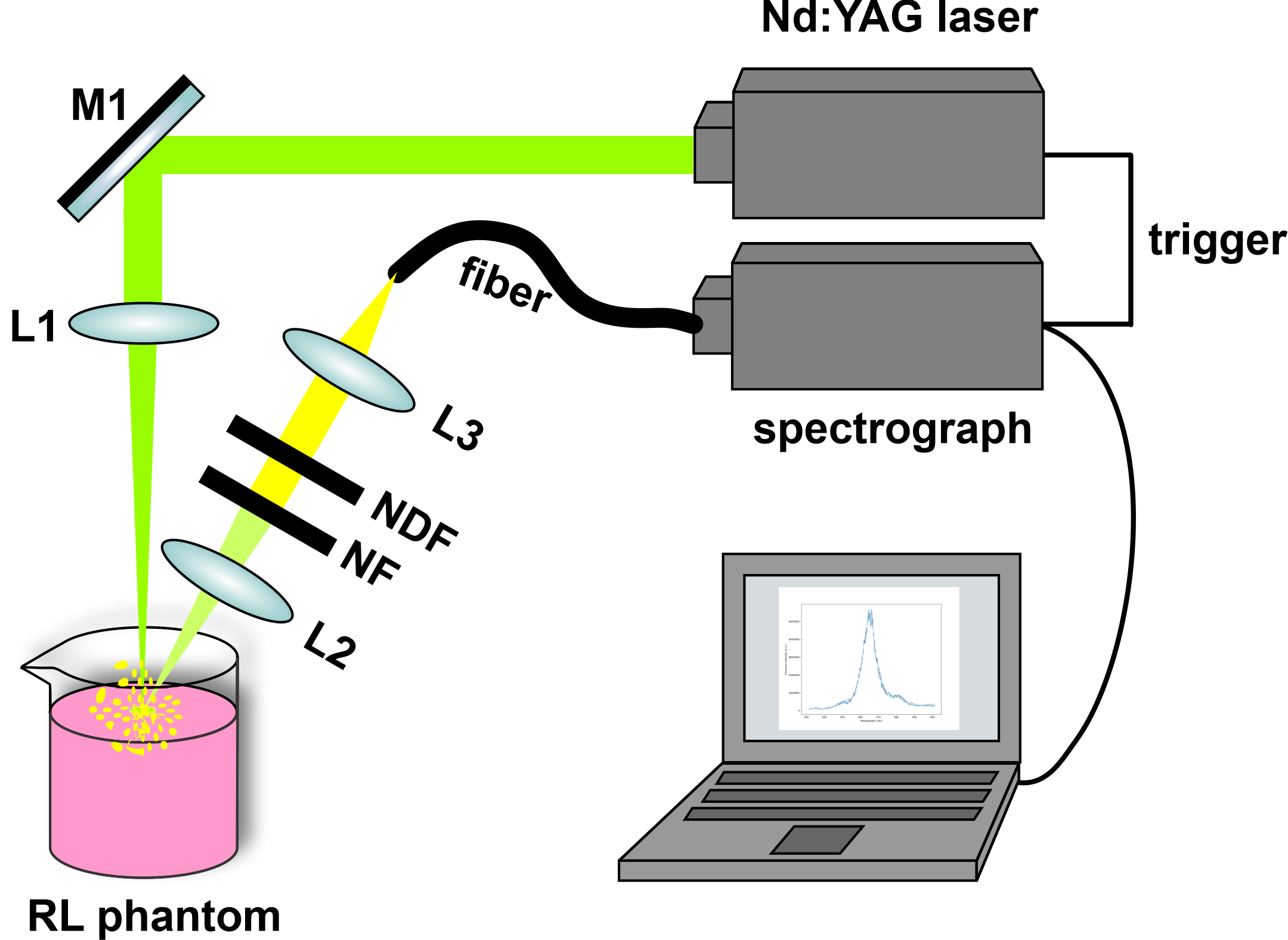}
		\caption{RL experimental setup from our previous study~\cite{Ni2023a}. The pump light from a pulsed Nd:YAG laser is focused onto the sample. The back-scattered emission light is collected from the excitation spot with an angle of 45$^{\circ}$, and then guided by the optical fiber to a spectrograph. M: mirror; L: lens; NF: notch filter; NDF: neutral density filter. } 
		\label{fig_setup}
	\end{figure}
	
	The experimental setup is shown in Figure~\ref{fig_setup}~\cite{Ni2023a}. The liquid sample with a volume of 75~mL was filled into a beaker, resulting in a cylindrical shape with a diameter of 50~mm and a height of 38~mm. The pump light from a pulsed Nd:YAG laser (Q-smart 450, Quantel) was focused onto the sample, and the backscattered light was collected and directed to a spectrograph (Mechelle Me5000 Echelle, Andor). For each measurement, a laser pulse with a wavelength of 532~nm and a pulse duration of 5~ns was generated, and was focused on the sample surface with a spot diameter of 0.25~mm. The emission light from the same position of the excitation spot was detected with an angle of 45$^{\circ}$. The spectrograph has a spectral range from 200 to 975~nm with a spectral resolving power ($\lambda$/$\Delta\lambda$) of 6,000, i.e. a spectral resolution of 0.1~nm at the wavelength of 600~nm. For each sample, the laser pump energy $E$ was varied from 1.82~mJ to 40.86~mJ in a trend of half-Gaussian distribution to probe both the non-lasing and lasing regime of the RL emission. Each measurement was repeated five times, and the averaged values were used. 
	
	\subsection{Optimization of scaling parameters $k_1$ and $k_2$}
	The optimization of the scaling parameters $k_1$ and $k_2$ is estimated by the convergence of the data points to the identical sensing curve. To assess the convergence, the $R^2$ value, which estimates the goodness of the fit, was applied as the optimization function. In practice, since the minimization algorithms are more commonly available than maximization algorithms, the minimum value of $1-R^2$ was evaluated instead of the maximum value of $R^2$. Three scales were chosen as the initial assumption for $k_1$ and $k_2$: 10, 1 and 0.1. These scales are comparable or smaller than the values of $\mu_s'$ and $\mu_a$ in Equation~\ref{eq_fitting}, so that the OPs rather than the scaling parameters of $k_1$ and $k_2$ dominate the diffuse reflectance. The tested values of $k_1$ and $k_2$ that lead to a minimum value of the objective function of $1-R^2$ were returned as the optimal values. 
	
	Two different optimization methods: direct search algorithm \textit{Nelder-Mead} and Markov chain Monte Carlo (MCMC) algorithm \textit{Metropolis-Hastings} were applied to avoid the method-induced optimization bias. The former one was implemented by using the Python package of the "scipy.optimize.minimize" with "Nelder-Mead" as the optimizer, $1-R^2$ as the objective function and $k_{1init}$, $k_{2init}$ as the initial assumption. In the latter one, a Gaussian distribution near the initial assumption $k$ was proposed ($k$ includes $k_1$ and $k_2$). A random point $k'$ from the proposed distribution was selected, and accepted if $p(k)- p(k') \geq \alpha$, where $p()$ denotes the objective function of $1-R^2$ and $\alpha$ denotes a uniform random number between 0 and 1. 
	
	\section{Results}
	\subsection{Data preprocessing }
\begin{figure}[h!]
		\centering
		\includegraphics[width=1\textwidth]{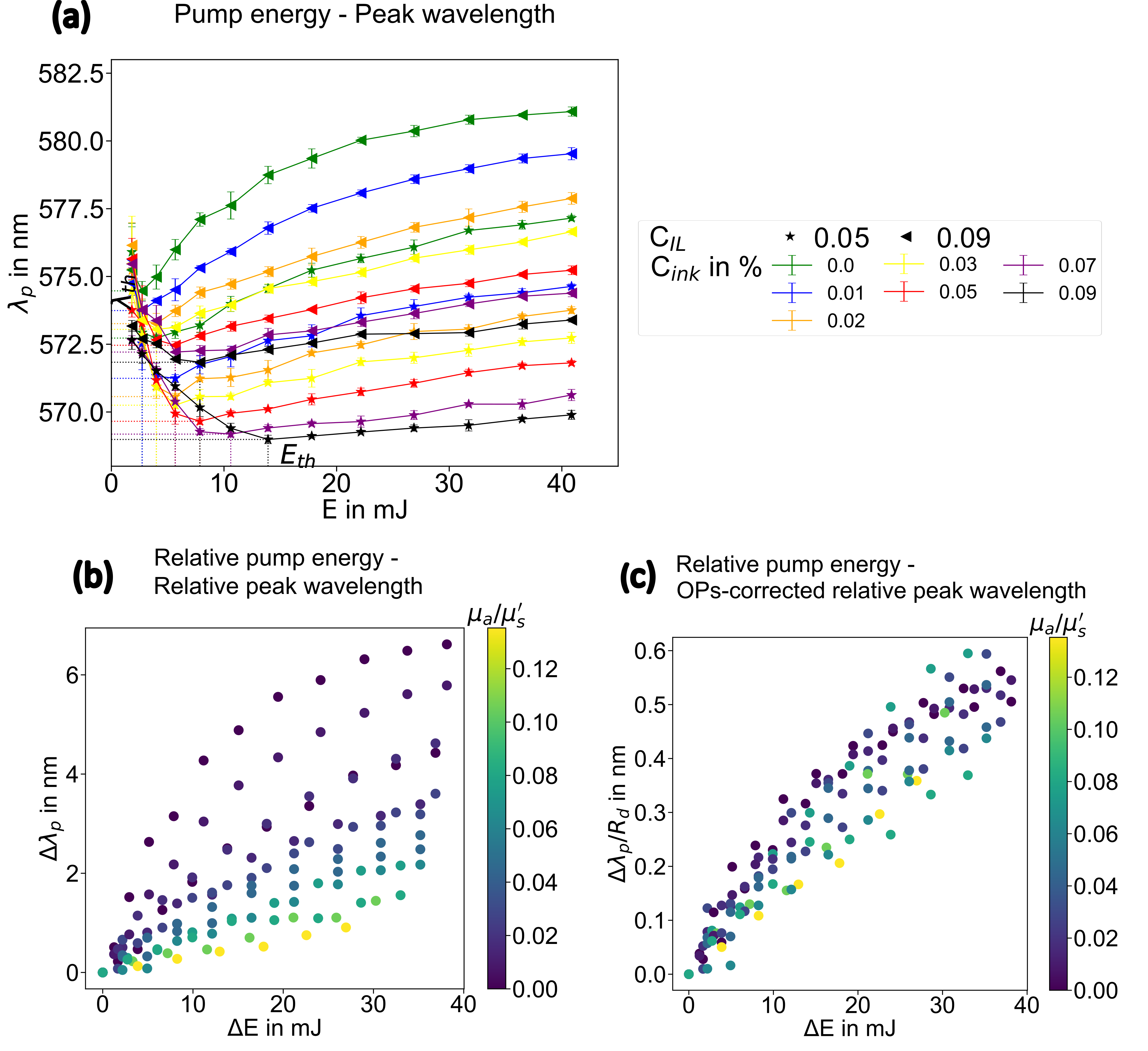}
		\caption{Data preprocessing for the DRS-RL sensing. (a) response of the RL peak wavelength to the pump energy that is applied to the samples. (b) preprocessed data in the lasing regime. Different colors represent the samples with different OPs. (c) data cloud after the correction of the OPs effect. The data points are converged when dividing by $R_d$, the value of which is calculated taking the arbitrary numbers of 8 and 2 for $k_1$ and $k_2$.} 
		\label{fig_preprocessing}
	\end{figure}

	The response of the RL peak wavelength $\lambda_p$ on the pump energy $E$ is shown in Figure~\ref{fig_preprocessing}~(a). For samples with the same IL concentration, increasing ink concentration induces the higher lasing threshold $E_{th}$ as well as the blueshift of the peak wavelength $\lambda_p$ in the lasing regime. For samples with the same ink concentration, increasing the IL concentration leads to an inverse response: a lower lasing threshold and a redshift of the peak wavelength in the lasing regime.  
	
	Figure~\ref{fig_preprocessing}~(b) shows the preprocessed data in the lasing regime. The peak wavelength of each RL sample with different OPs increases with increasing pump energy. The identical exponential response proves that the gain effect is identical and independent of the OPs. In Figure~\ref{fig_preprocessing}~(c), after correcting for the OPs effect by dividing the diffuse reflectance $R_d$, the data points representing the peak wavelength shift are remarkably converged. It is noteworthy that the $k_1$ and $k_2$ in $R_d$ take the arbitrary numbers of 8 and 2 to show the convergence. The convergence can be further improved by adjusting the scaling parameters of $k_1$ and $k_2$.
		
	\subsection{Model parameter optimization: $k_1$ and $k_2$}
	Table~\ref{tab_model_optimization} summarizes the optimization results of the two optimizers applying three different scales of initial assumptions of $k_1$ and $k_2$. Although the returned optimal values of $k_1$ and $k_2$ are different, the values of the fitting goodness $R^2$ are the same (the $R^2$ values differ only in the 9th digit after the decimal). The bias caused by the selection of the optimizers or initial assumptions is thus eliminated. Moreover, the optimal convergence of the data points to the sensing curve is 94\%, which is independent of the optimizer and initial assumption. 
	
	\begin{table}[!htb]
		\centering
		\caption{Optimization of the scaling parameters of $k_1$ and $k_2$. \label{tab_model_optimization}}
		\resizebox{1\textwidth}{!}{ 
			\begin{NiceTabular}{|c|cc|cc|cc|}		
				%\rowcolor{white}
				\CodeBefore
				\columncolor{gray!50}{3,5,7}
				\rowcolor{white!100}{1} 
				\Body
				\hline
				$k_{1init}$, $k_{2init}$
				& \multicolumn{2}{c|}{10, 10}
				& \multicolumn{2}{c|}{1, 1}
				& \multicolumn{2}{c|}{0.1, 0.1}\\
				\hline
				%\rowcolor{white}
				&Optimal $k_1$, $k_2$&$R^2$ 
				&Optimal $k_1$, $k_2$&$R^2$
				&Optimal $k_1$, $k_2$&$R^2$\\
				\hline
				\textit{Nelder-Mead}
				&4.22, 1.52&0.94 
				&0.43, 0.15&0.94
				&0.04, 0.02&0.94\\
				\hline
				\textit{Metropolis Hastings}
				&8.17, 2.94&0.94 
				&0.54, 0.19&0.94 
				&0.14, 0.05&0.94\\
				\hline
		\end{NiceTabular}}
	\end{table}

	\begin{figure}[h!]
	   \centering
	   \includegraphics[width=0.8\textwidth]{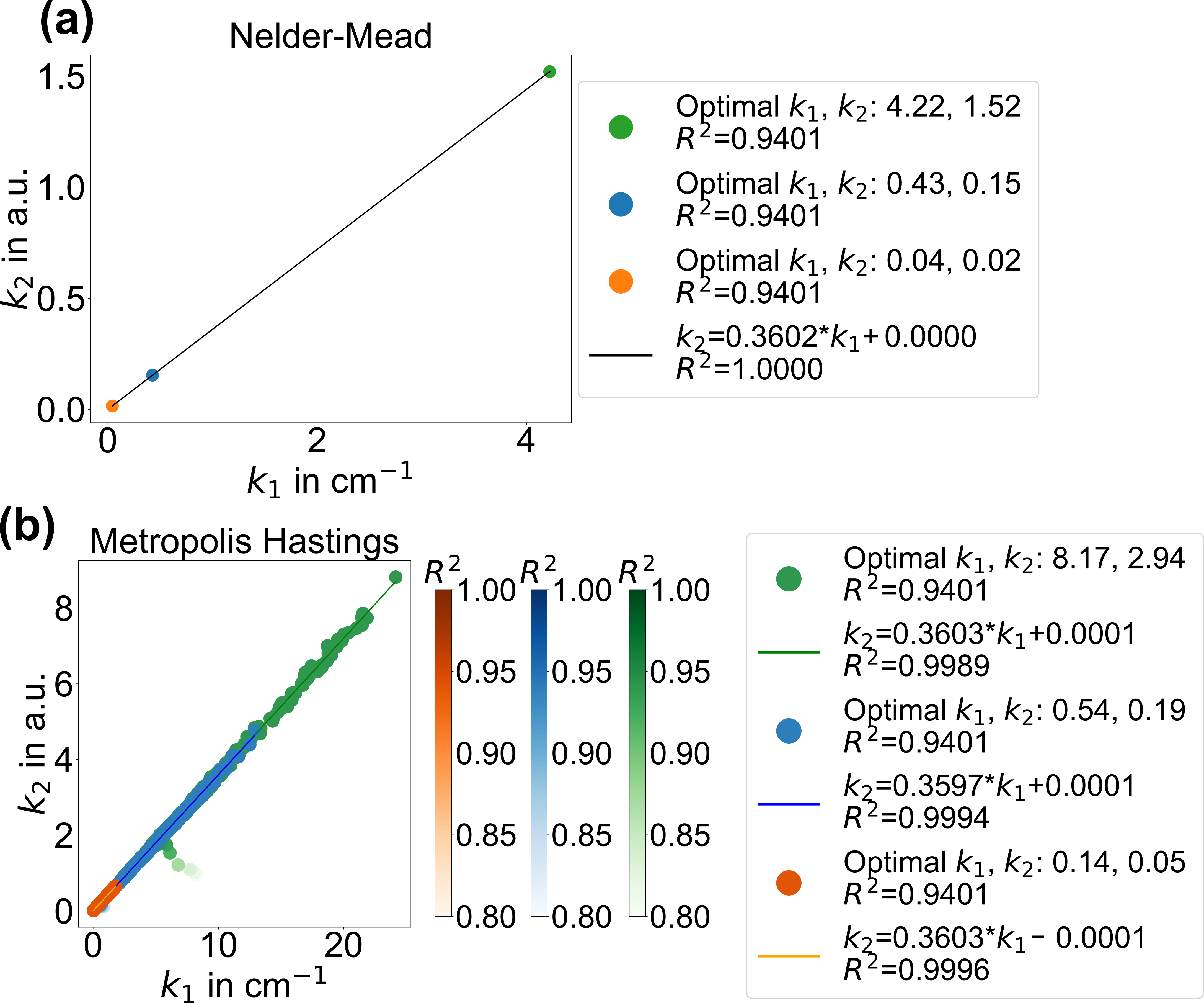}
	   \caption{Linearity of the $k_1$ and $k_2$ optimal values derived by the optimizer (a)     \textit{Nelder-Mead} and (b) \textit{Metropolis Hastings}.} 
	   \label{fig_parameter_optimization_comparison}
    \end{figure}	

	Furthermore, the relationship between the optimal $k_1$ and $k_2$ derived from both optimizers is illustrated in Figure~\ref{fig_parameter_optimization_comparison} for a comparison. The results from both cases reveal the same linearity between $k_1$ and $k_2$:
	\begin{equation}
		k_2=0.36 \cdot k_1.
		\label{eq_linearity}
	\end{equation}
	
	\subsection{DRS-RL sensing curve}
\begin{figure}[h!]
		\centering
		\includegraphics[width=1\textwidth]{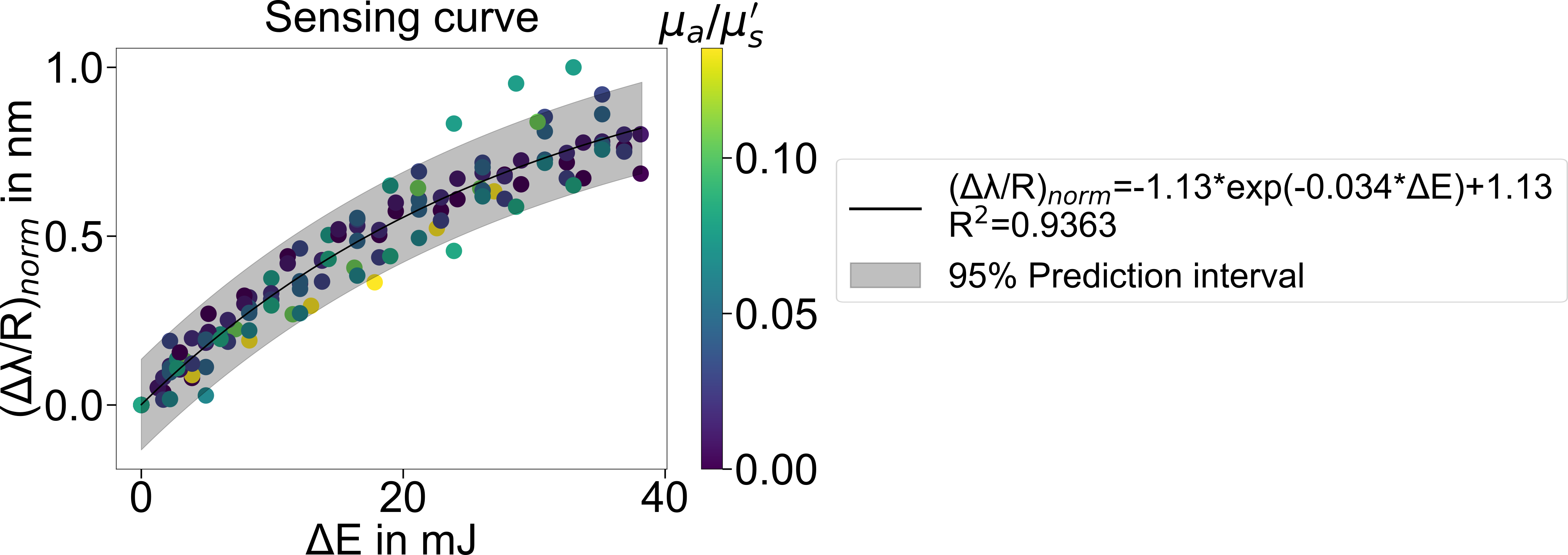}
		\caption{The normalized DRS-RL sensing curve. } 
		\label{fig_Sensing_curve}
	\end{figure}

	Figure~\ref{fig_Sensing_curve} shows the DRS-RL sensing curve after the normalization. The sensing data points fit well to the exponential sensing curve with $R^2 \approx 0.94$. The outlines located outside of the 95\% prediction interval may result from measurement errors. 
	
	The sensing curve is formulated by applying the optimal values of $k_1$ and $k_2$ in Table~\ref{tab_model_optimization} to the fitting function in Equation~\ref{eq_fitting}. Although the optimal values of $k_1$ and $k_2$ vary widely using different optimizers and different initial assumptions, a unique fitting function is returned after normalizing $\Delta \lambda_p / R_d$ to its maximum, as shown in Figure~\ref{fig_Sensing_curve}. Taking into account the Equation~\ref{eq_linearity}, the sensing curve is expressed as
	\begin{equation}
		\Delta \lambda_p \cdot \frac{k_1(1+0.36\mu_a)}{\mu_s'}|_{norm} = 1.13 \cdot (1-e^{-0.034 \cdot \Delta E}).
		\label{eq_sensing_curve1}
	\end{equation}
	A general scaling factor $k$ is further introduced to simplify the above equation:
	\begin{equation}
		k \cdot \frac{\Delta \lambda_p \cdot (1+0.36\mu_a)}{\mu_s'} = 1-e^{-0.034 \cdot \Delta E},
		\label{eq_sensing_curve2}
	\end{equation}
	$k = 9.59$ applying the measured ($\Delta\lambda_p$, $\Delta E$), the known ($\mu_s'$, $\mu_a$) values and the fitting parameters, and taking the average of the $k$ values calculated for each data point. Therefore, the mathematical expression of the sensing curve is finally formulated as:
	\begin{equation}
		9.59 \cdot \frac{\Delta \lambda_p \cdot (1+0.36\mu_a)}{\mu_s'} = 1-e^{-0.034 \cdot \Delta E}.
		\label{eq_sensing_curve3}
	\end{equation}
    
\begin{figure}[h!]
    	\centering
    	\includegraphics[width=0.7\textwidth]{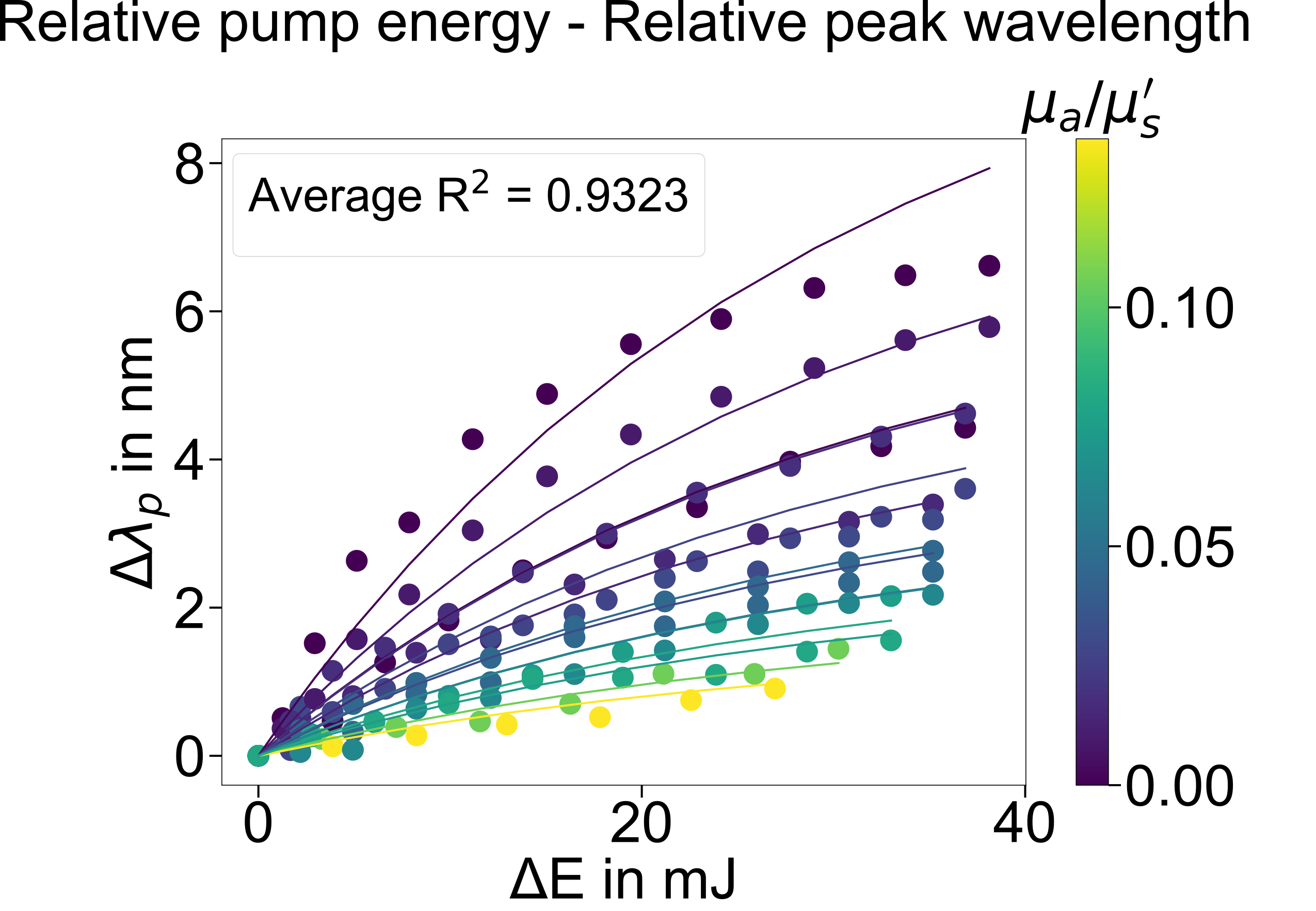}
    	\caption{The values of $\Delta \lambda_p$ calculated from the sensing curve or from the Equation~\ref{eq_sensing_curve} (represented by the curves), compared with the experimentally measured values of $\Delta \lambda_p$ (represented by the dots). } 
    	\label{fig_original_fit}
    \end{figure}

    From Equation~\ref{eq_sensing_curve3}, the RL sensing parameter of $\Delta \lambda_p$ is derived: 
    \begin{equation}
    	\Delta \lambda_p = \frac{1}{9.59} \cdot \frac{\mu_s'}{1+0.36\mu_a} \cdot (1-e^{-0.034 \cdot \Delta E}) . 
    	\label{eq_sensing_curve}
    \end{equation}
    This equation indicates that the RL peak wavelength shift in the lasing regime can be represented by three combined components: (1) a general scaling factor, (2) the OPs and (3) the gain effect induced by the pump energy. Furthermore, the calculated values of $\Delta \lambda_p$ from Equation~\ref{eq_sensing_curve} are approximate to the original values measured experimentally, with an average $R^2 = 0.9323$ as shown in Figure~\ref{fig_original_fit}. The deviation is larger for samples with lower values of $mu_a/\mu_s'$. One possible cause is the additional gain saturation as the RL medium has lower absorption loss and stronger feedback caused by the stronger scattering strength. Despite this deviation, the sensing curve and equation are representative for the experimental measurements and can be applied for optical sensing. 
    
	\subsection{Discussion and conclusion}
	Apart from the gain effect, RL in the backscattering direction responds similarly as the diffuse reflectance changes on optical properties. This study proposes a mathematical model of optical sensing that adapts the features of RLs to the common DRS model, the Zonios' model. A linear relationship between the two scaling parameters in the Zonios' model was revealed, and therefore the two scaling parameters can be simplified to one general scaling factor. In addition, it was shown that the pump energy induced gain effect in RL is universal and independent of the OPs of the samples investigated. The RL gain effect follows an exponential response of the transparent gain medium to the pump energy. For the first time, the OPs effect and the gain effect of the RL can be represented separately in a mathematical model for optical sensing. 
	
	The proposed optical sensing model was also experimentally validated. The reconstruction of the RL emission quantity using the sensing model can achieve an average $R^2$ of 93.23\%. However, the validation only included the turbid samples with two scattering coefficients. More samples with more scattering coefficients are needed to test the sensing model. Besides, the Zonios' model was used because it is simplified and also representative of diffuse reflectance. A limitation is that the phase function $p(\Theta)$ or the $g$ factor is not considered, although it has been shown that the $p(\Theta)$ or $g$ factor has an influence on the diffuse reflectance, especially when the source and detector separation is very close. Whether these factors also affect the RL quantities is still questionable. Therefore,  further investigations such as the influence of $p(\Theta)$ or $g$ factor, the influence of the setup geometry and their interrelationship are needed to complete the proposed DRS-RL sensing model. Furthermore, the transformation of the proposed sensing model from the liquid sample to a solid tissue phantom, which is inharmonious, needs to be investigated for a more practical application. 

	%\nolinenumbers
	
	\section*{Acknowledgment}
	This work was funded by the Deutsche Forschungsgemeinschaft (German Research Foundation – DFG 414732368). In addition, the authors gratefully acknowledge the funding of SAOT by the Bavarian State Ministry for Science and Art.
   
    \bibliography{library.bib}
    \bibliographystyle{unsrt}

\end{document}